\documentclass[twocolumn]{aastex631}
\usepackage[normalem]{ulem}
\usepackage{graphicx}
\usepackage{dcolumn}
\usepackage{bm}
\usepackage{color}
\usepackage{array}
\usepackage{multirow}
\usepackage{amsmath}
\usepackage{bm}
\usepackage{subfigure}
\usepackage{orcidlink}

\newcommand{\f}{\frac}

\newcommand{\n}{\nonumber}
\newcommand{\p}{\partial}

\renewcommand\arraystretch{1.3}

\begin{document}
\title[The reconstruction of dark energy with Ridge Regression Approach]{The reconstruction of dark energy with Ridge Regression Approach}

\author{Long Huang \orcidlink{0000-0003-4545-7066}}
 \affiliation{Xinjiang Astronomical Observatory, Chinese Academy of Sciences, Urumqi 830011, China}
     \affiliation{University of Chinese Academy of Sciences, Beijing, 10039, China}

 \author{Xiaofeng Yang$^{*}$}

 \affiliation{Xinjiang Astronomical Observatory, Chinese Academy of Sciences, Urumqi 830011, China}
  \affiliation{Key Laboratory of Radio Astronomy, Chinese Academy of Sciences, Nanjing 210008, China}
    \affiliation{Key Laboratory of Radio Astrophysics in Xinjiang Province, Urumqi 830011, China}

 \author{Xiang Liu}
 \affiliation{Xinjiang Astronomical Observatory, Chinese Academy of Sciences, Urumqi 830011, China}
  \affiliation{Key Laboratory of Radio Astronomy, Chinese Academy of Sciences, Nanjing 210008, China}
    \affiliation{Key Laboratory of Radio Astrophysics in Xinjiang Province, Urumqi 830011, China}

\begin{abstract}
It may be determined by non-parametric method if the dark energy evolves with time. In order to avoid instability of derivative for functional data, we linearize the luminosity distance integral formula in nearly flat space by adopting Lagrange interpolation for numerical integral, and propose a method of combining PCA and biased estimation on the basis of ridge regression analysis to reconstruct regression parameters. Meanwhile present an principal component selection criterion to better distinguish between $\Lambda CDM$ and $w(z) \ne  - 1$ models by reconstruction.  We define type I error is the situation that ${w_{true}} = -1$ but ${w _{recon}} \ne -1$, and type II error is the situation that ${w _{true}} \ne -1$ but ${w _{recon}} = -1$, and use the various $w(z)$ functions to test the method. The  preliminary test results demonstrate the PCA-biased method can be used to determine the most probable behavior of $w(z)$. Finally, we apply this method to recent supernova measurements,  reconstructing the continuous history of w(z) out to redshift z = 1.5.

\end{abstract}


\section{Introduction}

Since the cosmic acceleration was firstly discovered in 1998 \citep{riess1998}, physicists predict the existence of dark energy for explaining the accelerating expansion of the universe, on the other hand, the nature of dark energy whether or not evolves with time has also become a significant issue \citep{Linder}. The various observational data can be used to test the dark energy equation of state.

On the data aspects, astronomers are getting more and higher precision original observational data including Type Ia SNe \citep{Amanullah,Betoule,Scolnic}, Hubble parameter $H(z)$ \citep{Lewis}, cosmic microwave background radiation(CMB) \citep{Hu,Wmap,Planck2015,Planck2016}, and large scale structure (LSS) \citep{Eisenstein}. Although the correction methods for apparent magnitude of SNe Ia including SALT2 \citep{Guy2005,Guy2007}, SALT2 with the improved absolute magnitude depend on the prior cosmological models or the dark energy equation of state, the calibrated data can still be applied to partially recover the state parameter, $w(z)$.
Meanwhile Hubble parameters including some data obtained from the age-redshift relationship of galaxies
can be directly used to reconstruct the dark energy $w(z)$ models because its correction does not need to depend on a prior cosmological model \citep{Jimenez2002}. Therefore, original observation data including Type Ia SNe, Hubble parameter $H(z)$, cosmic microwave background radiation(CMB), and large scale structure (LSS) can be employed to obtain statistical results to measure the dynamic property of dark energy.

 Betoule1 et al. in 2014 \citep{Betoule} used a Joint Light-Curve Analysis of the SDSS-II and SNLS3 Supernova Survey and obtained the statistical result which indicates $w(z)$ model including linear and constant $w(z)$ has not evident superiority, comparing to $\Lambda CDM$ model by minimum chi-square. Even so, we still consider using non-parametric methods to examine whether or not $w(z)$ model has obvious advantage compared to $\Lambda CDM$.

 The non-parametric methods have several types: i) Principal Component Analysis (PCA), includes nonlinear PCA \citep{Huterer} and linear PCA \citep{Clarkson}, this method provide a good idea to test $w(z)$ model, the testing yields good result, it can recover $\Lambda CDM$ model, but it may be difficult to recover $w(z)$ model because of the instability of derivative. On the other hand, the selection of principal component has some artificial effects; ii) Gaussian Processes (GP) \citep{Holsclaw,Shafieloo,Seikel}, the advantage of GP method is that the calculation is simple, but the covariance function parameters have a significant impact on the calculation of results, and its selection has a certain arbitrariness. iii) PCA with the smoothness prior \citep{Crittenden2009,Crittenden2012,Zhao}, although the results obtained by this method are good, considering that it is a non-linear regression, the calculation is more complicated, and the choice of principal components and the covariance function parameters also have the above problems.

 We consider a method of combining numerical integral by Lagrangian interpolation, PCA analysis and biased estimation on the basis of ridge regression analysis, we apply this method to test the dark energy equation of state. At the same time, we propose a new principal component selection criterion to better distinguish between $\Lambda CDM$ and $w(z) \ne  - 1$ models by reconstruction.

\section[THE RECONSTRUCTION OF PARAMETER USING PCA WITH BIASED ESTIMATION]{THE RECONSTRUCTION OF PARAMETER USING PCA WITH BIASED ESTIMATION }
\subsection{Linearization of the relation between of luminosity distance and redshift}

We concern that the large positive and negative space curvature has not been obviously observed by observational data, and approximately adopt curvature ${\Omega _k} = 0$. Then to avoid the unstable problems of first and second derivation of luminosity distance, which are used to reconstruct $w(z)$ by second derivative formula of $w(z)$  \citep{Clarkson}. We use Gauss integration to linearize the luminosity distance formula, and adopt the simplest Lagrange's interpolation method to get a linear regression, which can be used to directly  reconstruct $w(z)$.

\textbf{The luminosity distance formula ${d_l}$ is
\begin{equation}\label{eq1}
{d_l} = \frac{{(1 + z)c}}{{{H_0}}}\int_0^z {\frac{{dz}}{{h(z)}}} ,
\end{equation}
where ${H_0}$ is the Hubble constant at redshift $z = 0$, $h(z) = H(z)/{H_0}$, $H(z)$ is the Hubble parameter at redshift $z$.}
We use Gauss integration to linearize the luminosity distance formula, the linear regression can be written as
\begin{equation}\label{eq2}
{d_H} - d_H^{prior} = y = X\beta ,
\end{equation}
which $\frac{{{d_H}}}{{{H_0}}} = {d_l}$, ${d_l}$ is the luminosity distance,  $\beta  = \Delta {(\frac{1}{h})^\prime } = d(\frac{{{H_0}}}{H} - \frac{{{H_0}}}{{{H_{prior}}}})/dr$, $r = {\raise0.7ex\hbox{$1$} \!\mathord{\left/
 {\vphantom {1 {1 + z}}}\right.\kern-\nulldelimiterspace}
\!\lower0.7ex\hbox{${1 + z}$}}$, and $X = {X'}T$, $T$ is Trapezoid numerical integration matrix, ${X'}$ is the coefficient matrix of Gauss integration, which is given by
\begin{equation}\label{eq3}
{X_{i{j^\prime }}} = \int_0^{{r_i}} {{L_j}(r) \cdot {r^{ - 2}}} dr ,
\end{equation}
which ${{L_i}(r)}$ is Lagrange multiplication operator function. When we choose $\Delta {(\frac{1}{h})'}$ as the interpolation points, because the boundary condition is ${\left. {\Delta (\frac{1}{h})} \right|_{r = 1}} = {\left. {(\frac{{{H_0}}}{H} - \frac{{{H_0}}}{{{H_{prior}}}})} \right|_{r = 1}} = 0$, it can be obtained by integral of $\Delta {(\frac{1}{h})'}$, and ${\left. {\int_0^r {K({r^{''}},{r'})} d{r^{''}}} \right|_{r = 1}} = 0$, which $K$ is covariance matrix, the boundary condition can be satisfied. We select the number of interpolation points $N \ge 20$, and use PCA with biased estimation to reconstruct regression parameter $\beta$.

\subsection{Biased estimation and PCA}

Considering a linear regression
\begin{equation}\label{eq4}
y = X\beta,
\end{equation}
assume both of $y$ and $\beta$ are function data, we put a zero mean Gaussian prior with covariance matrix ${K }$ on the regression coefficient $\beta$, the biased estimate of $\beta$ is  \textbf{\citep{Rasmussen}}
\begin{equation}\label{eq5}
{\hat \beta ^*} = K{(K + {C_{\hat \beta }})^{ - 1}}\hat \beta,
\end{equation}
which $\hat \beta $ is the unbiased estimate of $ \beta $, ${C_{\hat \beta }}$ is the covariance of $\hat \beta $, and ${C_{\hat \beta }} = {({X^T}{\Sigma _y}^{ - 1}X)^{ - 1}}$, $\Sigma _y$ is the covariance of data $y$, covariance function at $r$ and $r'$ written as

\begin{equation}\label{eq6}
k(r,r') = {\sigma ^2}*{e^{ - \frac{1}{2}{{(\frac{{\left| {r - r'} \right|}}{l})}^s}}}.
\end{equation}
where $\sigma ^2$ is the variance, $l$ is the correlation lengths, $k(r,r')$ is the element of matrix ${K }$. When $s = 2$, covariance matrix condition number will be too large to reduce the impact of input errors on estimated parameters. It is same for other forms of the $k(r, r')$ function, so we set the value of $s \in (1.5,{\kern 1pt} {\kern 1pt} 1.9)$. It can avoids some artificial effects for the choice of the  parameters $s$,  and also decrease matrix condition number, which can reduce the impact of input errors on estimated parameters, and avoid unstable problem in parameter values.

 The biased covariance of biased estimate of ${\hat \beta ^ * }$ is (see Appendix ~\ref{A5})
\begin{equation}\label{eq7}
{V_{{{\hat \beta }^*}}} = E[({\hat \beta ^*} - \beta )({\hat \beta ^*} - \beta )'] = Z{C_{\hat \beta }}{Z^T} + (Z - I)\beta {\beta ^T}{(Z - I)^T} ,
\end{equation}
where $Z = K{(K + {C_{\hat \beta }})^{ - 1}}$, $I$ is the unit matrix.\\

 Here we use the linear regression method that combined PCA with biased estimate based on ridge regression analysis \textbf{\citep{Frank,Wold}}, to decrease variance and reconstruct the biased estimate $\hat \beta $, the linear regression can be expressed as
\begin{equation}\label{eq8}
{{\hat \beta }^*} = {P^T}\alpha  = K({K^{ - 1}}{P^T})\alpha ,
\end{equation}
and
\begin{equation}\label{eq9}
{V_{{{\hat \beta }^*}}}= {P^T}\Lambda P.
\end{equation}
 where $\alpha$ is the regression coefficient vector, $P^T$ is the eigenmatrix, $\Lambda$ is the diagonalizable matrix of eigenvalue. Because ${P^T}\alpha$ is smooth estimate values, the right form of Eq. (\ref{eq8})  can satisfy that new continuous estimates are available, but does not change estimates. Using the orthonormality condition, the estimates ${\hat \alpha }$ can be computed as $\hat \alpha  = P{{\hat \beta }^*}$, its covariance matrix is diagonal, thus we can choose appropriate principal component to reconstruct the estimate $\hat \beta $.

\subsection{The choice of  principal component coefficient ${\hat \alpha_i}$ and covariance function parameters}

We assume a linear transformations $c = q\beta $ about linear regression $y = X\beta$, making sure the covariance matrix ${C_{\hat c}}$ of unbiased estimate of $\hat c$ is diagonal, then we put a zero mean Gaussian prior with covariance matrix ${K_c}$ on the parameter $c$, which ${K_c}$ is diagonal, and its element is ${k_{{c_i}}}^2$, minimizing the trace of Eq. (\ref{eq7}) $Tr({V_{{{\hat c}^ * }}})$ is equal to ${k_{{c_i}}} = \left| {{c_i}} \right|$, after that we combine Eq. (\ref{eq5}) and use biased estimate ${\hat c^ * }$ to substitute $c$, and find if ${k_{{c_i}}}$ has a real solution, ${\hat c }$ should satisfy condition ${\hat c_i}^2 \ge 4{\sigma ^2}({\hat c_i})$ (see Appendix ~\ref{B}). Thus, we give an interesting criteria which expressed as
\begin{equation}\label{eq10}
{\hat \alpha_i} = \left\{ \begin{array}{l}
{{\hat \alpha}_i}{\kern 1pt} {\kern 1pt} {\kern 1pt} {\kern 1pt} {\kern 1pt} {\kern 1pt} {\kern 1pt} {\kern 1pt} {\kern 1pt} {\kern 1pt} {\kern 1pt} (\left| {{{\hat \alpha}_i}} \right| \ge 2\sigma ({{\hat \alpha}_i}))\\
0{\kern 1pt} {\kern 1pt} {\kern 1pt} {\kern 1pt} {\kern 1pt} {\kern 1pt} {\kern 1pt} {\kern 1pt} {\kern 1pt} {\kern 1pt} {\kern 1pt} {\kern 1pt} {\kern 1pt} {\kern 1pt} {\kern 1pt} (\left| {{{\hat \alpha}_i}} \right| < 2\sigma ({{\hat \alpha}_i})).
\end{array} \right.
\end{equation}
We define type I error is the situation that ${\beta _{true}} = 0$ but ${\beta _{recon}} \ne 0$, and type II error is the situation that ${\beta _{true}} \ne 0$ but ${\beta _{recon}} = 0$. If we choose $\sigma ({\hat \alpha_i}) $ as criteria, which will increase the probability of making a type I error and decrease the probability of making a type II error, whereas if we choose $3\sigma ({\hat \alpha_i})  $ as criteria, which will increase the probability of making a type II error and decrease the probability of making a type I error.

If we assume $k(r,{r'}) = {\sigma ^2}$, minimizing $Tr({V_{{{\hat \beta }^ * }}})$ to be equal to ${\sigma ^2} \ge \beta _{\max }^2$, so we give $\sigma  = {\left| {{{\hat \beta }^ * }} \right|_{\max }}$, and the value of ${\left| {{{\hat \beta }^ * }} \right|_{\max }}$ trend to be stable with increasing of $\sigma$ (see Appendix ~\ref{B}). We select the value of $s$ in the range of $1.5 \le s \le 1.9$, which has little or not impact on the variety of ${\beta _{recon}}$, and the value of correlation lengths $l$ also has much less effect on ${\beta _{recon}}$, we will discuss this issue later on.\\

\subsection{The  reconstruction for $w(z)$}
We use the regression functional data $\Delta (1/h)$ and $\Delta {(1/h)'}$ to reconstruct $w(z)$, the reconstruction formula is
\begin{equation}\label{eq11}
{w_{recon}} = \frac{{2(1 + z){{(1/h)}'} + 3/h}}{{3\left\{ {{{(1 + z)}^3}{\Omega _{{m_0}}}{{(1/h)}^2} - 1} \right\}(1/h)}},
\end{equation}
which $(1/h) = \Delta (1/h) + {({H_0}/H)_{prior}}$, and ${(1/h)'} = \Delta {(1/h)'} + {({H_0}/H)_{prior}'}$.

Then we can take various $w(z)$ models to test the method.

\begin{figure}[tbp]
\centering
\includegraphics[scale=0.6,angle=0]{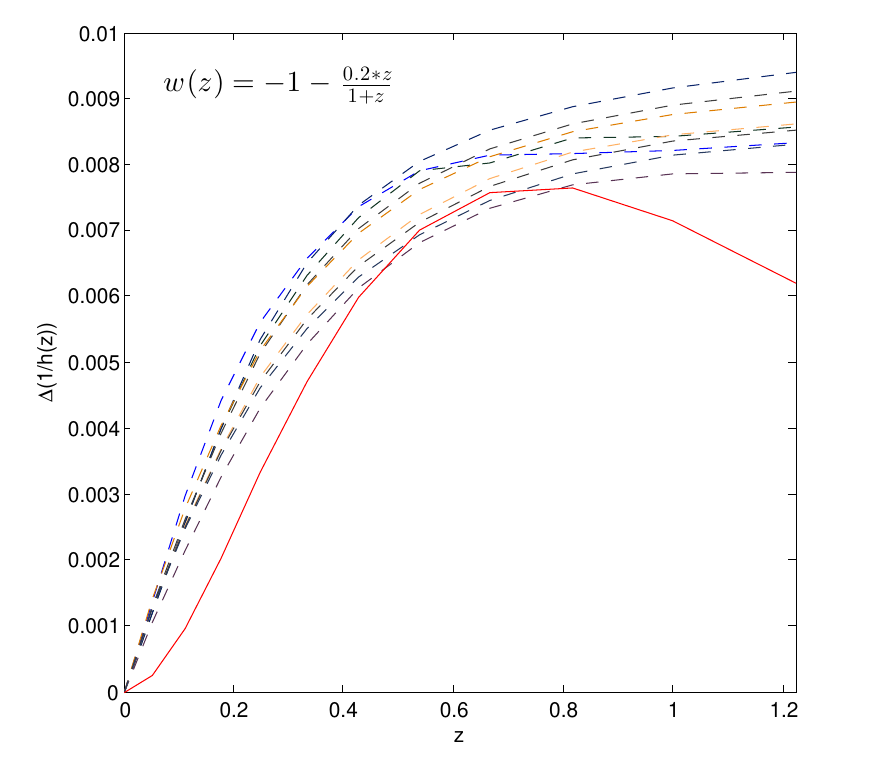}
\caption{The red line is the theoretical function used for sampling, and the dashed lines are the reconstruction results of $\Delta (\frac{1}{h})$  for correlation lengths l in the range of $0.15 \le l \le 0.5$. The results show that the value of l has not munch affect on the reconstruction results. }
\label{fig:detah}
\end{figure}

\section[TESTING THE METHOD BY THE VARIOUS $w(z)$ FUNCTIONS]{TESTING THE METHOD BY THE VARIOUS $w(z)$ FUNCTIONS }

We construct hypothetical ${d_H}$ data sampled from various $w(z)$ models, with the redshift in range of $0 < z \le 1.3$, and assume statistical uncertainty is ${\sigma _m} = 0.15mag$. The test $w(z)$ functions include the following forms:
\begin{equation}\label{eq12}
\begin{array}{l}
{\rm{w(z)  =    -  1}}\\
w(z) =  - 1 + \frac{{0.3 \cdot z}}{{1 + z}}\\
w(z) =  - 1 - 1.5ln(1 + z) \cdot \exp (\frac{{ - ln{{(1 + z)}^2}}}{{{{0.5}^2}}})\\
w(z) =  - 1 + 0.3\tanh (3 \cdot z){\kern 1pt} {\kern 1pt} {\kern 1pt} .
\end{array}
\end{equation}
 i) constant $w$ model; ii) the linear model which is often used in parametric method \citep{Linder,Maor,Riess2004,Riess2007,Sahni}; iii) the feature model \citep{Xia},where $ln$ is natural logarithm function; vi) the transition model which is reconstructed as fiducial model in nonparametric method \citep{Huterer,Corasaniti}. We use Eq. (\ref{eq2}) , (\ref{eq8}) and (\ref{eq11}) , and choose a flat $\Lambda CDM$ model as a prior model to obtain ${w_{recon}}$,  meanwhile we give the $\Delta {(\frac{1}{h})_{recon}}$ by the different values of correlation lengths $l$, their reconstructions are shown in Fig.~\ref{fig:detah}, which show $l$ has much less effect on reconstruction. The tests demonstrate that the selection of covariance function parameters $l$ and $s$ have not obvious influence on the calculations, and the value of $\sigma $ is given by explicit form from the analysis of the above section.

\begin{figure*}[htbp]
\includegraphics[scale=0.8]{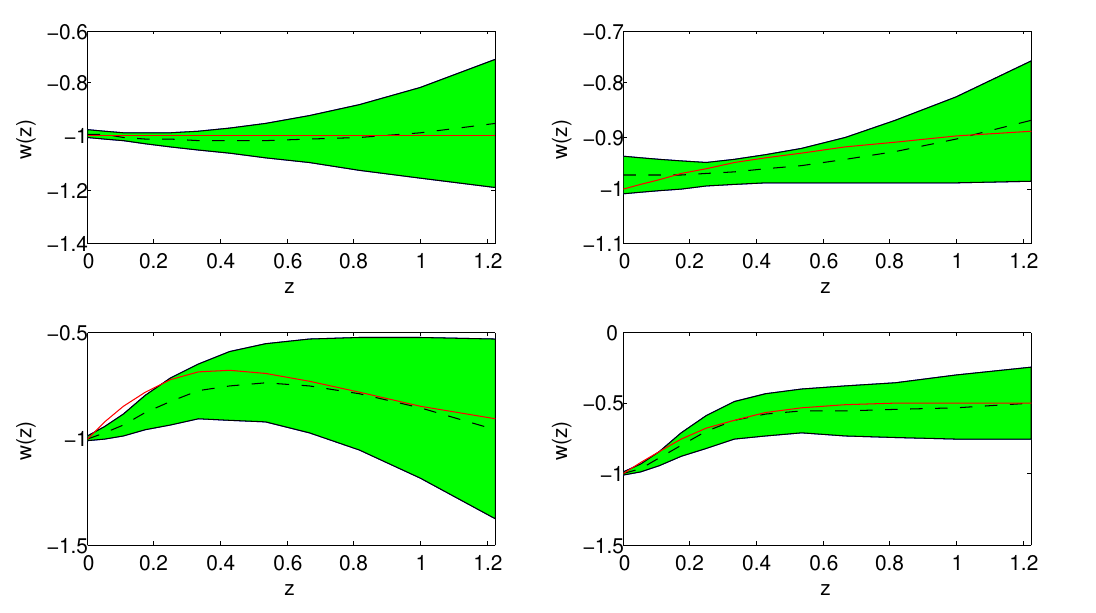}
\caption{The reconstruction results of $w(z)$ from ${d_H}$ data which are sampled from various $w(z)$ models together with the luminosity distance formula. The upper left figure is the reconstruction results for test models $w(z) =  - 1$ or $\Delta \overline w (z) = \overline {\left| {1 + w(z)} \right|}  <0.05$, the results show that when $\Delta \overline w (z) = \overline {\left| {1 + w(z)} \right|}  <= 0.05$ or $w=-1$, the reconstruction result is $w=-1$ if we choose $2\sigma ({\hat \alpha _i})$ as criteria (Eq. (\ref{eq10}) ), but the reconstruction result will have slightly deviated from $w =  - 1$ if we choose $\sigma ({\hat \alpha _i})$ as criteria. It shows that the method can be used to test models with slight deviations for $w(z) =  - 1$. The top right and bottom graphs are the reconstruction results for linear model, feature model, and transition model, it shows that our method has certain applicability to the recovery of ${\rm{w(z) }} \ne {\rm{  -  1}}$ models by same criteria .}
\label{fig:wz}
\end{figure*}

 The reconstructions of $w(z)$ and the functions are shown in Fig.~\ref{fig:wz}, the test results  indicate if $\Delta \overline w (z) = \overline {\left| {1 + w(z)} \right|}  <=0.05$  for various $w(z)$ models including $w(z) = -1$, the probability of making a type I error for ${w_{recon}} \ne 1$  as shown in the upper left figure is almost zero ($1\% $)  in the test which are not less 100 for the times of sampling; otherwise, if $\Delta \overline w (z) = \overline {\left| {1 + w(z)} \right|}  > 0.05$, the probability of making a type I error for ${w_{recon}} = 1$ is not more than $10\% $ as shown in other three graphs. We choose principal components by a new criteria, the goal is to better identify the highest priority for the $\Lambda CDM$ or $w(z) \ne  - 1$ models.  When $\Delta \overline w (z) = \overline {\left| {1 + w(z)} \right|}  <= 0.05$ or $w=-1$, the reconstruction results are $w=-1$ by our criteria (Eq. (\ref{eq10}) ), When $\Delta \overline w (z) = \overline {\left| {1 + w(z)} \right|}  >= 0.05$, the reconstruction results are $w \ne  - 1$ by same criteria.

\section[USING THE JLA SAMPLE TO RECONSTRUCT  $w(z)$]{USING THE JLA SAMPLE TO RECONSTRUCT  $w(z)$}

\subsection{Used data}

The SDSS-II and SNLS3 Supernova Survey Joint Light-Curve Analysis is called JLA for short. The Supernova Legacy Survey Program used a large CCD mosaic MegaPrime at the Canada-France-Hawaii Telescope to detect and monitor approximately 2000 high-redshift Supernovae between 2003 and 2008. 239 SNe Ia based on the first three years of data is contained in JLA sample, the goal of the Survey is to investigate the expansion history of the universe, improve the  constraint of cosmological parameters, as well as dark energy study, including the measure of the time-averaged equation of state of dark energy $w$ to 0.05 (statistical uncertainties only) in combination with other measurements and to 0.10 considering systematic effects \citep{Conley}.

 The SDSS-II Supernova Survey used the SDSS camera \citep{Gunn1998} on the SDSS 2.5 m telescope \citep{York,Gunn2006} at the Apache Point Observatory (APO) to search for SNe in the northern fall seasons (September 1 through November 30) of 2005 to 2007. Until running on the end of the year 2007, a wide variety of sources including solar system objects, galactic variable stars, active galactic nuclei, supernovae (SNe), and other astronomical transients were observed \citep{Sako2007}, 403 sources were identified as SNe \citep{Betoule}.

 In 2014 an large catalogue was released containing light curves, spectra, classifications, and ancillary data of 10,258 variable and transient sources by this Survey \citep{Sako2014}, The release resulted in the largest sample of supernova candidates ever compiled with 4607 likely supernovae, 500 of which have been confirmed as SNe Ia by the spectroscopic follow-up. JLA sample consists of a selection of 374 SNe Ia from this spectroscopic sample. The rest of JLA sample are taken from the C11 compilation, comprising SNe from SDSS, SNLS, HST and several nearby experiments \citep{Conley}. This extended sample of 740 SNe Ia is called the JLA sample.

\subsection{SALT2 calibration for JLA sample}
We consider the prior dark energy equation of state is unknown, so we use SALT2 and Taylor expansion of ${{\rm{d}}_H} - z$ relation to directly calibrate JLA sample, which can simplify the problem.

 The distance modulus ${\mu _{ob}}$ correction formula is given by SALT2 model \citep{Guy2005,Guy2007}
 \begin{equation}\label{eq13}
{\mu _{B,ob}} = {m_B} - {M_B}{\rm{ + }}\alpha  \times {x_1}{\rm{ + }}\beta  \times c,
\end{equation}
where ${m_B}$ corresponds to the observed peak magnitude in rest frame $B$ band, ${x_1}$ describes the time stretching of the light curve, $c$ describes the SN colour at maximum brightness, and $\alpha $, $\beta $ are nuisance parameters in the distance estimate. ${M_B}$ is the absolute B-band magnitude, which depends on the host galaxy properties \citep{Betoule}. Notice that ${M_B}$ is related to the host stellar mass (M stellar ) by a simple step function
 \begin{equation}\label{eq14}
{M_B} = \left\{ \begin{array}{l}
M_B^1{\kern 1pt} {\kern 1pt} {\kern 1pt} {\kern 1pt} {\kern 1pt} {\kern 1pt} {\kern 1pt} {\kern 1pt} {\kern 1pt} {\kern 1pt} {\kern 1pt} {\kern 1pt} {\kern 1pt} {\kern 1pt} {\kern 1pt} {\kern 1pt} {\kern 1pt} {\kern 1pt} {\kern 1pt} {\kern 1pt} {\kern 1pt} {\kern 1pt} {\kern 1pt} {\kern 1pt} {\kern 1pt} {\kern 1pt} {\kern 1pt} {\kern 1pt} {\kern 1pt} {\kern 1pt} {\kern 1pt} {\kern 1pt} {\kern 1pt} {\kern 1pt} {\kern 1pt} if{\kern 1pt} {\kern 1pt} {\kern 1pt} {M_{stellar}} < {10^{10}}{M_ \odot }\\
M_B^1 + {\Delta _M}{\kern 1pt} {\kern 1pt} {\kern 1pt} {\kern 1pt} {\kern 1pt} {\kern 1pt} {\kern 1pt} {\kern 1pt} otherwise
\end{array} \right.
\end{equation}
Here ${M_ \odot }$ is the mass of the Sun.

Meanwhile, the relation of distance modulus $\mu $ and luminosity distance ${d_H}$ can be written as
 \begin{equation}\label{eq15}
\mu  = 5{\log _{10}}{d_H} + 25 - 5{\log _{10}}{H_0}
\end{equation}

We use SALT2 and Taylor expansion of ${{\rm{d}}_H} - z$ relation to directly calibrate JLA sample, the Taylor expansion of ${{\rm{d}}_H} - z$ relation can be given by
 \begin{equation}\label{eq16}
{d_{H,th}} = \frac{c}{{1 - y}}\left\{ {y - \frac{{{q_0} - 1}}{2}{y^2} + \left[ {\frac{{3{q_0} - 2{q_0} - {j_0}}}{6} + \frac{{ - {\Omega _{{k_0}}} + 2}}{6}} \right]{y^3}} \right\},
\end{equation}
where $y = z/(1 + z)$. In order to reduce calculation error for high redshift data, we take this variable substitution. ${q_0}$  is the deceleration parameter, ${j_0}$ is the jerk parameters, and ${\Omega _{{k_0}}}$ is the curvature term.

The ${\chi ^2}$ of JLA data can be calculated as
 \begin{equation}\label{eq17}
{\chi ^2} = \Delta {d_H}^TC_{{d_H}}^{ - 1}\Delta {d_H},
\end{equation}
where $\Delta {d_H} = {d_{H,ob}} - {d_{H,th}}$. ${C_{{d_H}}}$ is the covariance matrix of ${d_H}$, which can be given by the propagation error formula together with ${C_\mu }$. ${C_\mu }$ is the covariance matrix of the distance modulus $\mu $, we only consider statistical error, and
 \begin{equation}\label{eq18}
\begin{array}{l}
{C_{\mu ,stat}} = {V_{{m_B}}} + {\alpha ^2}{V_{{x_1}}} + {\beta ^2}{V_c} + 2\alpha {V_{{m_B},{x_1}}} - 2\beta {V_{{m_{B,c}}}}\\
{\kern 1pt} {\kern 1pt} {\kern 1pt} {\kern 1pt} {\kern 1pt} {\kern 1pt} {\kern 1pt} {\kern 1pt} {\kern 1pt} {\kern 1pt} {\kern 1pt} {\kern 1pt} {\kern 1pt} {\kern 1pt} {\kern 1pt} {\kern 1pt} {\kern 1pt} {\kern 1pt} {\kern 1pt} {\kern 1pt} {\kern 1pt} {\kern 1pt} {\kern 1pt} {\kern 1pt} {\kern 1pt} {\kern 1pt} {\kern 1pt} {\kern 1pt} {\kern 1pt} {\kern 1pt}  - 2\alpha \beta {V_{{x_1},c}}
\end{array}
\end{equation}
From Eq. (\ref{eq17}), we can get the minimum chi-square without systematic errors is ${\chi ^2}/d.o.f. = 723/740$, which is consistent with the result obtained by calibrating the distance modulus ${\mu _B}$.

\subsection{Reconstruction results for the dark energy equation of state}

\begin{figure}[htbp]
\centering
\includegraphics[scale=0.6,angle=0]{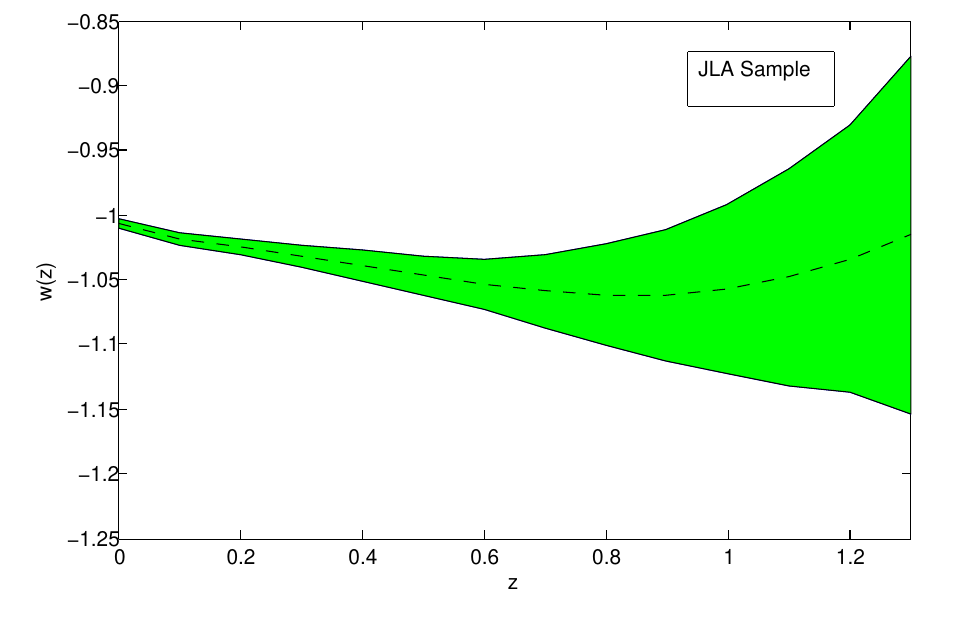}
\caption{The reconstruction results of $w(z)$  from JLA sample, which is calibrated by SALT2 model together with the kinematical model of universe.}
\label{fig:wjla}
\end{figure}

 We use JLA sample to reconstruct $w(z)$. Firstly, we combine the luminosity distance data and Eq. (\ref{eq2}) to derive the estimation of $\Delta {(\frac{1}{h})^\prime } $ , and then adopt Eq. (\ref{eq8})  to reconstruct $\Delta {(\frac{1}{h})^\prime }$, meanwhile employ integration to obtain $\Delta (\frac{1}{h}) $, in order to reduce the probability of making a type II error, we choose criteria $\sigma ({\hat \alpha_i})$ and $2\sigma ({\hat \alpha_i})$ for Eq. (\ref{eq10}) . Finally, we adopt second derivative formula Eq. (\ref{eq11}) to obtain ${w_{recon}}$. The reconstruction is shown in Fig.~\ref{fig:wjla}, the results indicates the equation of state has not obviously deviated from $w =  - 1$, or it may be consistent with $\Lambda CDM$.\\

\section{CONCLUSIONS}

We have presented a new, nonparametric reconstruction technique, which combines PCA, biased estimation, and numerical integral by Lagrangian interpolation techniques, this new method provide a good idea for solving the derivative problems of large variances functional data. The advantages of this method are as follows: i) adopting Lagrangian interpolation to avoid instability of derivative for functional data, ii) presenting an interesting principal component selection criterion to better identify the highest priority for the $\Lambda CDM$ or $w(z) \ne  - 1$ models, meanwhile the choice of prior does not have much impact on reconstruction results, iii) using linearization formula to make the calculation easier. The test results demonstrate the PCA-biased method can be used to determine the most probable behavior of w(z) and to infer how likely a target trajectory is given the current data. Thus it can be used to accept or reject classes of $\Lambda CDM$ model.

 We employ this method for the dark energy equation of state and applied it to JLA supernova sample, the results are consistent with $\Lambda CDM$. In the future, we hope to observe more high redshift Type Ia SNe data $z \ge 1$, it will be more convenient to reconstruct the dark energy equation of state.

 On the other hand, although the origin and composition properties of dark energy remain unknown, we still can use the observational data to test the dynamic property of dark energy. A solid measurement that $w = -1{\kern 1pt} {\kern 1pt} {\kern 1pt}$ or $\ne -1$ (which would rule out the cosmological constant) would have profound implications for cosmology and particle physics, it may bring us more new physical ideas.

\appendix

\section{ The biased estimate and biased covariance \label{A}}

Considering a linear regression $Y = X\beta $, the covariance of data $Y$ is $\Sigma _y$, which is diagonal matrix, then adopt least square method to derive the biased estimate  of parameter $\beta $
\begin{equation}
\hat \beta  = {({X^T}{\Sigma _y}^{ - 1}X)^{ - 1}}{X^T}{\Sigma _y}^{ - 1}y ,
\end{equation}
and its covariance
\begin{equation}
{C_{\hat \beta }} = {({X^T}{\Sigma _y}^{ - 1}X)^{ - 1}} .
\end{equation}
we put a zero mean Gaussian prior with covariance matrix $K$ on the parameter $\beta$, then the residual sums of squares can be written as
\begin{equation}
\phi (\beta ) = {(\hat \beta  - \beta )'}C_{\hat \beta }^{ - 1}(\hat \beta  - \beta ) + {\beta '}{K^{ - 1}}\beta ,
\end{equation}
Minimize $\phi (\beta )$ and get
\begin{equation}
{\hat \beta ^*} = K{(K + {C_{\hat \beta }})^{ - 1}}\hat \beta  = Z\hat \beta ,
\end{equation}
and its biased covariance
\begin{equation}
{V_{{{\hat \beta }^*}}} = E[({\hat \beta ^*} - \beta )({\hat \beta ^*} - \beta )'] = Z{C_{\hat \beta }}{Z^T} + (Z - I)\beta {\beta ^T}{(Z - I)^T} .
\label{A5}
\end{equation}
where $Z = K{(K + {C_{\hat \beta }})^{ - 1}}$, $I$ is the unit matrix.\\

\section{ The ridge regression analysis for biased covariance \label{B}}

Making a linear transformations for the regression coefficient
\begin{equation}
c = q\beta ,
\end{equation}
which ${C_{\hat \beta }} = {q^T}\Lambda q$, $q^T$ is the eigenmatrix, $\Lambda$ is the diagonalizable matrix of eigenvalue, then the covariance of unbiased estimate $\hat c$ is
\begin{equation}
{C_{\hat c}} = \Lambda .
\end{equation}
We put a zero mean Gaussian prior with covariance matrix ${{K_c}}$ on the parameter $c$, which ${{K_c}}$ is diagonal, and its element is $k_{{c_i}}^2$, then get the trace of biased covariance
\begin{equation}
Tr[{V_{{{\hat c}^ * }}}] = \sum\limits_i {\frac{{{\lambda _i}{{({K_c})}^2} \cdot {\lambda _i}(\Lambda )}}{{{{({\lambda _i}({K_c}) + {\lambda _i}(\Lambda ))}^2}}}}  + \sum\limits_i {\frac{{{c_i}^2{\lambda _i}{{(\Lambda )}^2}}}{{{{({\lambda _i}({K_c}) + {\lambda _i}(\Lambda ))}^2}}}} ,
\end{equation}
where ${V_{{{\hat c}^ * }}}$ is the biased covariance of biased estimate ${\hat c^ * }$, ${\lambda _i}({K_c})$ and ${\lambda _i}(\Lambda )$ are the eigenvalue of matrix ${{K_c}}$ and $\Lambda$. Minimize $Tr[{V_{{{\hat c}^ * }}}]$, and obtain
\begin{equation}
{k_{{c_i}}} = \left| {{c_i}} \right|.
\end{equation}
which ${c_i}$ is the element of regression parameter vector $c$.
If we use biased estimate ${\hat c^ * }$ to substitute $c$, and ${k_{{c_i}}}$ has a real solution, $\hat c$ should satisfy condition
\begin{equation}
{\hat c_i}^2 \ge 4{\sigma ^2}({\hat c_i}).
\end{equation}
where ${\hat c_i}$ is the element of unbiased estimate $\hat c$, ${\sigma ^2}({\hat c_i})$ is the element of matrix $\Lambda$.\\

\section*{Acknowledgments}

 We thank Dr. Ming Zhang for useful discussions. This work was supported by CAS Light of West China Program (No.2016-QNXZ-B-25), Xiaofeng Yang's Xinjiang Tianchi Bairen project and CAS Pioneer Hundred Talents Program.


\begin{thebibliography}{99}

\bibitem[\protect\citeauthoryear{Ade et al.}{2016}]{Planck2015}
Ade P. et al., 2016,  Astron. Astrophys.   594, A13 .

\bibitem[\protect\citeauthoryear{Aghanim et al.}{2016}]{Planck2016}
Aghanim N. et al., 2016, Astron. Astrophys.     596, A107.

\bibitem[\protect\citeauthoryear{Amanullah et al.}{2010}]{Amanullah}
Amanullah R. et al., 2010, Astrophys. J.    716,  712.

\bibitem[\protect\citeauthoryear{Betoule et al.}{2014}]{Betoule}
Betoule M. et al., 2014, Astron. Astrophys.     568,  A22.

\bibitem[\protect\citeauthoryear{Clarkson et al.}{2010}]{Clarkson}
Clarkson  C., Zunckel C., 2010, Phys. Rev. Lett.   104, 211301.

\bibitem[\protect\citeauthoryear{Conley et al.}{1998}]{Conley}
Conley  A. et al., 2010, Astrophys. J. Suppl. S.   192, 1.

\bibitem[\protect\citeauthoryear{Corasaniti et al.}{1998}]{Corasaniti}
Corasaniti P. S.,  Copeland E. J.,  2003,  Phys. Rev. D    67 , 063521 .

\bibitem[\protect\citeauthoryear{Riess et al.}{2009}]{Crittenden2009}
Crittenden R. G.  et al.,  2009, JCAP    0912, 025.

\bibitem[\protect\citeauthoryear{Crittenden et al.}{2012}]{Crittenden2012}
Crittenden  R. G.  et al., 2012, JCAP    1202, 048.

\bibitem[\protect\citeauthoryear{Eisenstein et al.}{2005}]{Eisenstein}
Eisenstein D. J. et al., 2005,  Astrophys. J.   633,  560.

\bibitem[\protect\citeauthoryear{Frank et al.}{1993}]{Frank}
Frank L. E., Jerome H. F., 1993, Technometrics, 35(2), 109.

\bibitem[\protect\citeauthoryear{Gunn  et al.}{1998}]{Gunn2006}
Gunn J. E. et al., 2006,  Astron. J.   131, 2332.

\bibitem[\protect\citeauthoryear{Gunn et al.}{1998}]{Gunn1998}
Gunn J. E. et al., 1998, Astron. J.   116, 3040.

\bibitem[\protect\citeauthoryear{Guy et al.}{2005}]{Guy2005}
Guy J. et al., 2005, Astron. Astrophys.   443, 781.

\bibitem[\protect\citeauthoryear{Guy et al.}{2007}]{Guy2007}
Guy J. et al., 2007, Astron. Astrophys.   466, 11.

\bibitem[\protect\citeauthoryear{Holsclaw  et al.}{2010}]{Holsclaw}
Holsclaw  T. et al.,  2010, Phys. Rev. Lett.    105, 241302.

\bibitem[\protect\citeauthoryear{Jimenez et al.}{2003}]{Huterer}
Huterer D., Starkman G., 2003, Phys. Rev. Lett.    90, 031301.

\bibitem[\protect\citeauthoryear{Hu et al.}{2002}]{Hu}
Hu W.,  Scott D., 2002,  Annu. Rev. Astron. Astr.   40, 171-216.

\bibitem[\protect\citeauthoryear{Riess et al.}{2002}]{Jimenez2002}
Jimenez R., Loeb A., 2002, Astrophys. J.   573, 37.

\bibitem[\protect\citeauthoryear{Komatsu et al.}{2011}]{Wmap}
Komatsu E. et al., 2011, Astrophys. J. Suppl. S.   192, 18.

\bibitem[\protect\citeauthoryear{Lewis et al.}{2002}]{Lewis}
Lewis A., Bridle S., 2002, Phys. Rev. D    66,  103511.

\bibitem[\protect\citeauthoryear{Linder et al.}{2003}]{Linder}
Linder E. V., 2003,  Phys. Rev. Lett.   90,  091301.



\bibitem[\protect\citeauthoryear{Maor  et al.}{2002}]{Maor}
Maor I. et al., 2002, Phys. Rev. D    65, 123003.

\bibitem[\protect\citeauthoryear{Rasmussen et al.}{2006}]{Rasmussen}
Rasmussen C. E. , Williams C. K. I., 2006, the MIT Press.

\bibitem[\protect\citeauthoryear{Riess et al.}{1998}]{riess1998}
Riess A. G. et al., 1998, Astrophys. J.   116, 1009.

\bibitem[\protect\citeauthoryear{Riess et al.}{2004}]{Riess2004}
Riess A. G. et al.,  2004, Astrophys. J.    607, 665.

\bibitem[\protect\citeauthoryear{Riess et al.}{2007}]{Riess2007}
Riess  A. G.  et al., 2007, Astrophys. J.   659, 98.

\bibitem[\protect\citeauthoryear{Sahni et al.}{2006}]{Sahni}
Sahni  V.,  Alexei S., 2006,  Int. J. Mod. Phys. D.    15, 2105.

\bibitem[\protect\citeauthoryear{Sako et al.}{1998}]{Sako2007}
Sako  M. et al., 2007,  Astron. J.   135, 348.

\bibitem[\protect\citeauthoryear{Sako et al.}{2014}]{Sako2014}
Sako  M. et al. 2014, Publ. Astron. Soc. Pac.    130, 064002.

\bibitem[\protect\citeauthoryear{Scolnic et al.}{2018}]{Scolnic}
Scolnic D. M. et al., 2018,  Astrophys. J.    859,  101.

\bibitem[\protect\citeauthoryear{Seikel et al.}{2012}]{Seikel}
Seikel M. et al.,  2012,  JCAP    06, 036.

\bibitem[\protect\citeauthoryear{Shafieloo et al.}{2012}]{Shafieloo}
Shafieloo A. et al., 2012, Phys. Rev. D   85, 123530.

\bibitem[\protect\citeauthoryear{Wold et al.}{2001}]{Wold}
Wold S., Sjostrom M., and Eriksson L., 2001, Chemometrics and intelligent laboratory systems, 58(2), 109.

\bibitem[\protect\citeauthoryear{Xia et al.}{1998}]{Xia}
Xia J. Q. et al.,  2006,  Phys. Rev. D   74, 083521 .

\bibitem[\protect\citeauthoryear{York et al.}{1998}]{York}
York  D. G. et al., 2000,  Astron. J.   120, 1579.

\bibitem[\protect\citeauthoryear{Zhao et al.}{2012}]{Zhao}
Zhao  G. B. et al.,  2012,  Phys. Rev. Lett.    109, 171301 .


\end{thebibliography}
\end{document}